\documentclass[prl,aps,amssymb]{revtex4}
\usepackage{graphicx}
\usepackage{rotating}

\newcommand{\nc}{\newcommand}
\nc{\bb}{\begin{equation}} \nc{\bba}{\begin{eqnarray}}
\nc{\bbas}{\begin{eqnarray*}} \nc{\ee}{\end{equation}}
\nc{\eea}{\end{eqnarray}} \nc{\eeas}{\end{eqnarray*}}
\nc{\1}{1\!\!1} \nc{\R}{I\!\!R}
\nc{\vecna}{\mbox{\boldmath $\nabla$}} \nc{\ug}{\; = \;}
\nc{\Abf}{\mbox{\boldmath $A$}} \nc{\Jbf}{\mbox{\boldmath $J$}}
\nc{\Vbf}{\mbox{\boldmath $V$}} \nc{\ebf}{\mbox{\boldmath $e$}}
\nc{\jbf}{\mbox{\boldmath $j$}} \nc{\rbf}{\mbox{\boldmath $r$}}
\nc{\sbf}{\mbox{\boldmath $s$}} \nc{\vbf}{\mbox{\boldmath $v$}}
\nc{\xbf}{\mbox{\boldmath $\xi$}} \nc{\nabf}{\mbox{\boldmath $\nabla$}}
\nc{\Hbf}{\mbox{\boldmath $H$}} \nc{\sigbf}{\mbox{\boldmath $\sigma$}}
\nc{\imp}{\mbox{\boldmath $p$}}
\nc{\Dc}{{\cal D}} \nc{\Ec}{{\cal E}} \nc{\Fc}{{\cal F}} \nc{\Gc}{{\cal G}}
\nc{\Lc}{{\cal L}} \nc{\Pc}{{\cal P}} \nc{\Rc}{{\cal R}} \nc{\Sc}{{\cal S}}
\nc{\Uc}{{\cal U}} \nc{\Vc}{{\cal V}} \nc{\Wc}{{\cal W}} \nc{\erm}{{\rm e}}
\nc{\ebfs}{\mbox{\boldmath $\scriptstyle e$}}
\nc{\rbfs}{\mbox{\boldmath $\scriptstyle r$}}
\nc{\vbfs}{\mbox{\boldmath $\scriptstyle v$}}
\nc{\rot}{{\widetilde{\rho}}} \nc{\Pt}{{\widetilde{\Pc}}}
\nc{\rz}{\rbf_0} \nc{\rt}{(\rbf,t)} \nc{\rtt}{(\rbf(t),t)} \nc{\rit}{(\rbf_i,t)}
\nc{\rto}{(\rbf,t_0)} \nc{\rzs}{\rbfs_0} \nc{\rts}{(\rbfs,t)} \nc{\rtts}{(\rbfs(t),t)}
\nc{\lai}{{\lambda_i}} \nc{\Lai}{{\Lambda_i}} \nc{\launo}{{\lambda_1}}
\nc{\Launo}{{\Lambda_1}} \nc{\tz}{t_0} \nc{\pa}{\partial}
\nc{\eps}{\varepsilon} \nc{\lap}{\triangle}

\begin{document}


\title{Quantum Lyapunov exponents}

\author{ P. Falsaperla and G. Fonte }
\address{ Dipartimento di Fisica e Astronomia, Universit\`a di Catania,
          Corso Italia 57, I-95129 Catania, Italy           \\
          and Istituto Nazionale di Fisica Nucleare, Sezione di Catania,
          Corso Italia 57, I-95129 Catania, Italy}

\author{ Giovanni Salesi }
\address{ Facolt\`a di Ingegneria, Universit\`a Statale di Bergamo,
          I-24044 Dalmine (BG), Italy \\
          and Istituto Nazionale di Fisica Nucleare, Sezione di Milano,
          I-20133 Milano, Italy}

\date{\today}

\begin{abstract}
We show that it is possible to associate univocally with each given solution
of the time-dependent Schr\"odinger equation a particular phase flow (``quantum
flow'') of a non-autonomous dynamical system. This fact allows us to introduce
a definition of chaos in quantum dynamics (quantum chaos), which is based on
the classical theory of chaos in dynamical systems. In such a way we can
introduce quantities which may be appelled {\em quantum Lyapunov exponents}.
Our approach applies to a non-relativistic quantum-mechanical system of $n$
charged particles; in the present work numerical calculations are performed
only for the hydrogen atom. In the computation of the trajectories we first
neglect the spin contribution to chaos, then we consider the spin effects
in quantum chaos. We show how the quantum Lyapunov exponents can be evaluated
and give several numerical results which describe some properties found in the
present approach. Although the system is very simple and the classical
counterpart is regular, the most non-stationary solutions of the
corresponding Schr\"odinger equation are {\em chaotic} according to our
definition.
\end{abstract}


\narrowtext

\maketitle

\section{Introduction}

Around the turn of the century, Poincar\'e pointed out a property which is
typical of most mechanical systems. Although the evolution is deterministic,
it shows an extreme sensitivity to initial conditions. Errors in these latter
can grow up at an exponential rate, and thus in pratice, long-time
predictions become impossible and the evolution appears irregular and random.
In the modern literature this phenonenon is called ``deterministic chaos''or
also ``classical chaos''.~\footnote{In the present paper we prefer the latter
denomination.}
In the context of classical mechanics, it upsets the laplacian ideal of exact
predictability, but, in the context of quantum mechanics, it rises even more
profound problems. Indeed, in virtue of Bohr's correspondence principle, one
expects that quantum mechanics must include or imitate, in some way, classical
chaos. The direct analogue of this  in quantum mechanics should be a sort
of exponential sensitivity to the initial conditions in the solutions of the
time-dependent Schr\"odinger equation.
The difficulty to put in evidence such a phenomenon is well
known \cite{Jensen,Hilbor} and the reason is obvious.
In quantum mechanics, instead of individual trajectories, we have a field:
the wave-function $\psi\rt$, whose time-evolution is given by
\bb
  \psi\rt = \Uc(t,\tz) \psi\rto,
  \label{eq1}
\ee
where $\Uc(t,\tz)$ is a unitary operator \cite{Reed} on $L^2$.
Thus, according to $L^2$-norm, two nearby initial wave-functions
remain close as time evolves. Owing to this fact, which seems to rule out any
possibility of a chaotic behaviour in quantum mechanics formally analogous to
the classical one, the investigations on the subject have moved toward a
different direction, that is in search of traces of classical chaos which
persist in the quantum world. Although this research has
obtained (see Ref.\ \cite{Jensen,Hilbor} and references therein)
interesting and far-reaching results,
it looks less fundamental than the attempt of finding out a property of the
quantum-mechanical time-evolution which is  analogous to classical chaos.
In the present
paper, we show that such a property really exists.
To put in evidence it, we give up, obviously, the too ``rough''
$L^2$-norm to measure wave-function modifications
and go to examine, so to say,``submicroscopical'' details
in the time-evolution~(\ref{eq1}).
In short, these details are the ``trajectories'' $\rbf=\rbf(t)$ determined
by the condition of being tangent at any point $\rbf$  to the
``velocity field'' $\Vbf\rt=\Jbf\rt/\rho\rt$ at time $t$ and at point
$\rbf$, where $\Jbf\rt$ and $\rho\rt$ are, respectively, the probability
density current and the probability density relative to the
wave-function $\psi\rt$.
We shall see that the set of these trajectories, distributed with the
probability density $\rho\rt$, can be considered as a particular phase
flow (quantum flow) of an usual non-autonomous dynamical system,
univocally determined by the wave-function $\psi\rt$ in~(\ref{eq1}).
This latter fact permits us to introduce a definition of chaos
for the time-evolution~(\ref{eq1}), founded only on the basic
concepts of the theory of chaos in dynamical systems.
We call such a definition of chaos {\em quantum chaos\/}.
As we shall see, our characterization of chaos
in quantum dynamics implies also a sort of
exponential sensitivity to the initial conditions in the Schr\"odinger equation
and thus, it is the direct analogue of classical chaos.
Notice that the emergence of quantum chaos, by ``unraveling'' the quantal
evolution~(\ref{eq1}) into ``individual trajectories'' is, in a way,
similar to the emergence of classical chaos in Halmitonian systems,
by ``unraveling'' the time-evolution of the phase space probability
distribution $\rho(q,p,t)$ into individual particle trajectories.
Indeed, since the equation (Liouville equation) satisfied by $\rho(q,p,t)$
is linear, slight modifications in the initial phase space
probability distribution never lead \cite{Hilbor}
to exponential divergences of $\rho(q,p,t)$,
although all or part of its ``underlying'' trajectories may be chaotic.

It should be observed that our approach is formally similar to the
one recently proposed~\cite{Holland,Parmenter,Faisal} within the
framework of the de Broglie-Bohm causal interpretation of
quantum mechanics (Bohmian mechanics), however it presents
the following novelties:
\begin{itemize}
\item[-] Our approach seems to be fuller.
\item[-] We do not make recourse to any new interpretation of
         quantum mechanics.
\item[-] We introduce suitable parameters (quantum Lyapunov exponents) to
         quantify quantum chaos.
\item[-] We show a calculation of these parameters in a system simple
         but realistic (hydrogen atom).
\item[-] We show intriguing relation\-ships between quantum-mechanical
         quantities along a given trajectory and the Lyapunov exponents
         of this latter.
\end{itemize}
Although all the numerical results presented (Sec.~5) concern the hydrogen
atom, our characterization of quantum chaos and the relative implications
(Secs.~2-4) are discussed in the general setting of a quantum system of $n$
charged particles.

\section{The Quantum Flow}

In this section we show how it is possible
to associate univocally with a given solution
$\psi\rt$ of the initial value problem (IVP) for the Schr\"odinger equation a
suitable phase flow  of an usual non-autonomous dynamical system.

Let us consider a system of $n$ particles with masses $m_1,\ldots,m_n$ and
charges $e_1,\ldots,e_n$.
The corresponding Schr\"odinger equation writes
\bb
  i\hbar\,\frac{\pa\psi\rt}{\pa t} =
     \left[-\sum_{i=1}^n\frac{\hbar^2}{2m_i}\,\lap_i +
     \sum_{i<j}\frac{e_ie_j}{|\rbf_i-\rbf_j|}\right]\psi\rt\,,
  \label{eq2}
\ee
where $\rbf=(\rbf_1,\ldots,\rbf_n)$, $\rbf_i\in\R^3,\ i=1,2,\ldots,n$.
Each wave-function
\bb
  \psi\rt=[\rho\rt]^{1/2}\,\erm^{iS(\rbfs,t)/\hbar}
  \label{eq2bis}
\ee
satisfying Eq.~(\ref{eq2}) generates a probability density current
\[
  \Jbf\rt = (\jbf_1\rt,\ldots,\jbf_n\rt),
\]
where
\bb
  \jbf_i\rt := \frac{\rho\rt}{m_i}\,\left[\nabf_iS\rt\right],
   \qquad i=1,2,\ldots,n,  \label{eq3}
\ee
and hence a  ``velocity field''
\bb
  \Vbf\rt = (\vbf_1\rt,\ldots,\vbf_n\rt)\,,
  \label{defV} 
\ee
where
\bb
  \vbf_i\rt := \frac{\jbf_i\rt}{\rho\rt},
  \qquad i=1,2,\ldots,n.
  \label{defvi} 
\ee

The vector field~(\ref{defV}) is not defined at any point
$\rt\in\R^{3n}\times\R_t$, where $\R_t$ denotes the time axis.
In general, we can say that it is defined on some set
\[
  \Pt = \R^{3n}\!\times\R_t\,/\,\Sc\,,
\]
where
\bbas
  \Sc=\{\rt\in\R^{3n}\times\R_t:
      &&\mbox{either~}\psi\rt=0\mbox{~(nodes)}\\
      &&\mbox{or~}    \psi\rt \mbox{~is not differentiable}\}.
\eeas
We call the set $\Pt$ {\em extended phase space\/}.
We also need  to consider the set
\[
  \Pc=\{\rbf\in\R^{3n}: \rt\in\Pt, \mbox{~for some~} t\}
\]
and the set
\[
  \Pc(t)=\{\rbf\in\R^{3n}:\rt\in\Pt, \mbox{~for~} t \mbox{~fixed}\}.\]
We call the sets
$\Pc$ and $\Pc(t)$ {\em phase space\/} and {\em phase space at  $t$\/},
respectively.
Let us now introduce in the extended phase space $\Pt$ the ``direction field''
($\Vbf\rt, 1$). It is easily seen~\cite{Arnold,Molenaar}
that the curves in $\Pt$,
which at any point $\rt$ have tangent directed as ($\Vbf\rt,1$) in that point,
can be described by the parametric
equations $\rbf=\rbf(t), t=t$, where $\rbf(t)$ satisfies the equation
\bb
\dot{\rbf}(t) = \Vbf\rt.
\label{rteq}
\ee
We name such curves {\em integral curves\/}.
Notice that Eq.~(\ref{rteq}) defines what in the modern literature
on chaos theory is referred to as {\em non-autonomous dynamical system\/}.
For all we have said above, this system is determined univocally by the solution $\psi\rt$
of a given IVP for the Schr\"odinger equation
\bb
\left\{
  \begin{array}{l}
    i\hbar \pa_t \psi\rt = H \psi\rt \\
    \psi(\rbf,0) =: \psi_0(\rbf), \qquad \psi_0(\rbf)\in\Dc(H),
                                  \quad \|\psi_0(\rbf)\|=1,
  \end{array}
\right.
\label{eq9}
\ee
where $H$ is the Hamiltonian operator in (\ref{eq2}) and $\Dc(H)$ its domain.
In association with the IVP~(\ref{eq9}) we may also consider the following IVP
\bb
\left\{
  \begin{array}{l}
    \dot{\rbf}(t) = \Vbf\rt \\
    \rbf(0) =: \rz, \qquad \rz\in\Pc(0).
  \end{array}
\right.
\label{eq10}
\ee

The solution of~(\ref{eq10}), when represented in
the extended phase space $\Pt$,
coincides with the integral curve which goes through the point $(\rz,0)$.
However, we prefer to think  of the solution of~(\ref{eq10})
as the path (orbit) in the phase space $\Pc$ followed, as time develops,
by a point starting from $\rz$ at $t=0$.
According to this representation, the solution of~(\ref{eq10})
will be called {\em trajectory\/}.
 Thus, the phase space $\Pc$ is the region (which can also coincide
 with the whole $\R^{3n}$)
accessible to our trajectories and the phase space at $t$,
$\Pc(t)$, is the subregion of $\Pc$
crossed by our trajectories at time $t$.~\footnote{According to this picture,
the point $\rbf$ when it varies on $\Pc(t)$ will be denoted by $\rbf(t)$.
The volume element $d\rbf$ on $\Pc(t)$ will be denoted by $d\rbf(t)$.}
It should be observed that both the trajectory $\rbf=\rbf(t)$ and the
corresponding orbit are tangent at any point $\rbf$ to the velocity field
$\Vbf\rt$ at time $t$ and at point $\rbf$ and that the orbit is
the projection onto $\Pc$ of the integral curve.
Obviously, in the spirit of the usual interpretation of quantum mechanics,
the trajectories of the IVP~(\ref{eq10}) do not correspond
to an effective motion of quantum particles.
However, this fact does not have any relevance at all
in our characterization of chaos in quantum dynamics.
Indeed, as we shall see, what is most important to this aim
is that the trajectories of the IVP~(\ref{eq10}) are {\em formally analogous\/}
to trajectories of real point particles.
Another consideration about our trajectories is that they must exist
uniquely and globally in time, i.e.\ for all $t\in\R_t$.
This property is, in fact, of basic importance to apply the theory of
chaos in dynamical systems.
Thus, we turn now our attention to it.
First of all, we note that a prerequisite of this property is that the
velocity field~(\ref{defV}) exists uniquely and globally in time, i.e.\ the
IVP~(\ref{eq9}) admits an unique and global solution.
In order that it be true, all is required is~\cite{Reed}  the
self-adjointness of $H$, property which in our case is valid~\cite{Reed}.
Consider now the main question:
does the IVP~(\ref{eq10}) determined by~(\ref{eq9}) admit an unique and
global solution?
This question turns out to be more difficult, because the solution
of~(\ref{eq10}) may run into catastrophic events, such as reaching
in finite time a point of the set $\Sc$ (the singularities of the
velocity field).
The question, which is also of basic importance in Bohmian mechanics,
has been investigated recently~\cite{Zanghi}.
The findings of the paper~\cite{Zanghi} are that, for a large class
of potentials (including the Coulomb potential in~(\ref{eq2}))
and under certain hypotheses (see~\cite{Zanghi} for details),
the IVP~(\ref{eq10}) admits an unique and global solution.
This basic property will be assumed throughout the present paper.
Now, in order to make more precise the definition of trajectory as well as
the definition of quantum flow to be given below,
we find it convenient to introduce the following mappings:
\begin{itemize}
\item[a)]
  $\Phi(\cdot,t_1,\rbf_1):\R_t \to \Pc$.
  This mapping represents the trajectory of the IVP~(\ref{eq10}),
  which goes through the point $\rbf_1$ at time $t_1$.
  Thus $\Phi(t,t_1,\rbf_1)=\rbf(t)$ represents the coordinate $\rbf$
  at time $t$ of the point pursuing this trajectory.
\item[b)]
  $\Phi(\cdot,t_1,\cdot):\R_t\times\Pc(t_1) \to \Pc$.
  In the theory of dynamical systems, this mapping is called {\em phase flow\/}.
  In our picture, it represents the set of all trajectories of
  the IVP(\ref{eq10}), passing through  $\Pc(t_1)$ at time $t_1$.
\item[c)]
  $\Phi(t,t_1,\cdot):\Pc(t_1) \to \Pc(t)$.
  We call this mapping {\em phase flow at $t$\/}.
\end{itemize}

The existence of these mappings is due~\cite{Arnold,Scheck} to the assumption
that the IVP~(\ref{eq10}) admits an unique and global solution.
This assumption implies also~\cite{Arnold,Scheck} the following properties:
\begin{itemize}
\item[d)]
  For  $t_1,t_2\in\R_t$ fixed, the mapping
  $\Phi(t_2,t_1,\cdot):\Pc(t_1) \to \Pc(t_2)$
  is one-to-one, differentiable and with
  inverse differentiable (diffeormophism).
\item[e)]
  $\Phi(t,t,\cdot)=I$ (identity).
\item[f)]
  $\Phi(t,t_2,\Phi(t_2,t_1,\rbf_1))=\Phi(t,t_1,\rbf_1)$
  (chain rule)~\footnote{In what follows, we set $t_1=0$.Thus $\rbf_1=\rz$.
To simplify notation, we shall write $\Phi(t,\rz),\Phi(\cdot,\rz)$ etc.,
for $\Phi(t,0,\rz),\Phi(\cdot,0,\rz)$ etc., respectively.}
\end{itemize}

Now, let us introduce on $\Pc(0)$ a statistical ensemble $\Ec$ of initial
conditions (points), distributed according to the probability density
$\rot_0(\rbf(0)) = \psi^\star_0(\rbf(0)) \psi_0(\rbf(0))=\rho_0(\rbf(0))$,
where $\psi_0(\rbf)$ is the wave-function appearing in the IVP~(\ref{eq9}),
and let us consider the time-evolution of these points according to
the IVP~(\ref{eq10}), i.e.\ the phase flow $\Phi(\cdot,\cdot)$ defined
on $\R_t\times\Ec$.
We call this special realization of the phase flow {\em quantum flow\/} (QF).
Thus from one hand, the QF is formally  similar to the phase flow of
an ordinary dynamical system, on the other hand it has characteristic
peculiarities.
In fact, the velocity field $\Vbf\rt$ in~(\ref{eq10}) depends
on a specific solution of the IVP~(\ref{eq9}) and the initial conditions are
distributed with a probability density, which is not uniform as in the phase
flow of dynamical systems, but is given by $\rho_0(\rbf(0))$.
The fact that to each given solution $\psi\rt$ of~(\ref{eq9})
there corresponds an unique dynamical system~(\ref{rteq})
and hence an unique QF, gives us the opportunity of defining a
chaotic behaviour in quantum dynamics.
Indeed, we may say that the given $\psi\rt$ is chaotic when the corresponding
QF presents trajectories which are chaotic, according to the
usual definition of classical chaos.
We call this characterization of chaos in quantum dynamics {\em quantum chaos\/}.

As we shall show in Sec.~5, our definition of quantum chaos implies also a
kind of extreme sensitivity to initial conditions, in the sense that a tiny
modification in $L^2-$norm in the initial condition $\psi(\rbf,0)$
in~(\ref{eq9}) gives rise to a dramatic modification in the QF pattern:
each chaotic trajectory turns into another diverging exponentially from it.
Strictly speaking, the definition of quantum chaos given as above
is not fully rigorous and does not put in evidence the role of the
ensemble $\Ec$.
A rigorous and satisfactory definition of quantum chaos as well as a
precise measure of it are postponed until Sec.~4, because first we need
to describe some properties of the QF and also to recall the methods
to quantify classical chaos.

\section{Some Properties of the Quantum Flow}

The properties of the QF are essentially those of an ordinary phase
flow. However, due to peculiarity of the QF, some of them show themselves under
new light and imply intriguing connections with quantum mechanical properties.
Thus, besides mentioning the main
properties of the QF, we dwell particularly upon these new aspects, giving
also some proof, when this turns out to be necessary to make the paper
clear and self-contained~\footnote{In what follows, we denote by $\psi\rt$ the
solution of the IVP~(\ref{eq9}) which gives rise to the QF considered.}.\\

{\bf Property 1.}
{\sl
Denoting by $\rot(\rbf(t),t)$ the probability density of points
$\rbf(t)$ in the ensemble $\Ec$ on $\Pc(t)$,
the assumption, in the previous section,
that the density of points in $\Ec$ at $t=0$ on $\Pc(0)$  is
$\rot_0(\rbf(0)) = \rho_0(\rbf(0))$,
implies~\cite{Holland,Zanghi}
\ $\rot\rtt=\psi^\star\rtt\psi\rtt=\rho\rtt$, \ $\forall t>0$
and $\forall \rbf(t) \in \Pc(t)$.
}\\

{\bf Property 2.}
{\sl
Let $\rz$ be a given point of $\Pc(0)$ and $d\rz$ be the volume
element around this point~\footnote{Throughout this paper {\em volume element}
means a volume small enough, in principle infinitesimal.}.
Let $\rbf(t)$ and $d\rbf(t)$ be the corresponding quantity tranformed
by the QF at $t>0$, i.e.\ $\rbf(t)=\Phi(t,\rz)$,
$d\rbf(t)=\Phi(t,d\rz)$.
Then, the probability measure $\rho\rtt\,d\rbf(t)$ is invariant under the QF,
i.e.
\bb
    \rho_0(\rz)\,d\rz = \rho\rtt\,d\rbf(t), \qquad\forall\,t>0.
    \label{invar}
\ee
}\\

{\sl Proof.} Think of the QF at $t$ (cf. points c) and d) in Sec.~2)
as the coordinate transformation
\bb
\Phi(t,\rbf(0)) = \rbf(t)
\label{phi}
\ee
from $\Pc(0)$ to $\Pc(t)$, $t>0$.
We have
\bb
d\rbf(t) = J(t,\rz)\,d\rz\,,
\label{eq13}
\ee
where $J(t,\rz)$ denotes the Jacobian at $\rz$~\footnote{The Jacobian depends
also on the initial time $t=0$. But according to footnote 3,
we have written $J(t,\rz)$ instead of $J(t,0,\rz)$.} of the
transformation~(\ref{phi}).
This Jacobian satisfies~\cite{Cauley} the equation
\bb
  \dot{J} = \left(\sum_{i=1}^n\,\nabf_i\cdot\vbf_i\right)\,J\,,
  \label{eq14}
\ee
where the quantities $\vbf_i$ are given by~(\ref{defvi}).
Integrate~(\ref{eq14}) with the initial condition $J(0,\rz)=I$,
and integrate the continuity equation for $\rho$ in the
Lagrangian form, i.e.,
\bb
\frac{d\rho}{dt} +
\rho \left(\sum_{i=1}^n\,\nabf_i\cdot\vbf_i\right) = 0\,,
\label{eq15}
\ee
with the initial condition $\rho_0(\rz)$.
We find
\[
\rho\rtt J(t,\rz) = \rho_0(\rz),\qquad \forall t>0,
\]
which, in virtue of~(\ref{eq13}),
yields the relationship~(\ref{invar}).$\square$\\

{\bf Property 3.}
{\sl
The growth rate under the QF of the size of the volume element  $d\rbf(t)$
around the point $\rbf(t)$ is given by
\bb
\frac{d\:d\rbf(t)}{dt} =
d\rbf(t)\left(\sum_{i=1}^n\,\nabf_i\cdot\vbf_i\rtt\right),\qquad
\forall\,t>0\,.
\label{eq16}
\ee
}\\

The proof is an easy consequence of Property 2 and of~(\ref{eq15}).
The integration of Eq.~(\ref{eq16}) over a finite volume
$\Vc(t)\subset\Pc(t)$ yields
\[
\frac{d \Vc(t)}{dt} =
\int_{\Vc(t)}\,\left(\sum_{i=1}^n\,\nabf_i\cdot\vbf_i\rtt\right)\,d\rbf(t).
\]
Since (see Eqs.~(\ref{eq2bis},\ref{eq3},\ref{defvi}))
\[
\sum_{i=1}^n\,\nabf_i\cdot\vbf_i\rtt =
\sum_{i=1}^n\,\frac{1}{m_i}\,\left(\lap_i S\rtt\right),
\]
we find that in a given region $\Omega\subset\Pt$, the QF is
{\em conservative\/} or {\em nonconsevative\/}, according to whether the value of
$\sum_{i=1}^n\,\frac{1}{m_i}\,(\lap_i S\rtt)$
in $\Omega$ is equal to zero or different from zero, respectively.

We have to recall now the basic elements to define and quantify chaos
in dynamical systems, i.e.\ the  definition and the main properties  of the
Lyapunov exponents (LEs) of a given trajectory.
We do that going in some details, because these are necessary to put
in evidence the interesting connections mentioned  at the beginning of
this section as well as to explain part of the numerical
results shown in Sec.~5.

According to Refs.~\cite{Benettin,Ruelle}, the LEs of the trajectory
$\Phi(\cdot,\rz)$ starting from
$\rz$ at $t=0$ and belonging to our QF,
are defined by the following limit
\bb
\lambda(\rz,\xbf_0) :=
\lim_{t\to+\infty}\,\frac{1}{t}\,\ln\|U_{\rzs}(t)\,\xbf_0\|\,,
\label{L00}
\ee
as function of $\xbf_0$, where $\|\cdot\|$ is the Euclidean norm,
$\xbf_0$ is an arbitrary starting vector of $\R^{3n}$, $\|\xbf_0\|=1$,
and $U_{\rbfs_0}(t)$ is an $3n\times 3n$
invertible matrix (flow matrix).
This matrix is the solution of the IVP
\bb
\left\{
  \begin{array}{l}
    \dot{U}_{\rbfs_0}(t) =
    \left\{
       \left.
          \frac{\pa \vbf_i}{\pa \rbfs_j}
       \right|_{\Phi(\cdot,\rz)}
    \right\}_{i,j=1}^{n} U_{\rbfs_0}(t)\\
    U_{\rbfs_0}(0) = I,
  \end{array}
\right.
\label{eq18}
\ee
obtained by linearization of~(\ref{eq10}).
Rigorous results on the existence of the limit~(\ref{L00})
and on the properties of the corresponding LEs can be
found in Refs.\ \cite{Benettin,Ruelle}
and references therein.
It is not easy to show (it would be also beyond the scope of
the present paper) that the findings in~\cite{Benettin,Ruelle},
proved under suitable hypotheses, can be taken over to the
dynamical system~(\ref{rteq}).
However, on the basis of a strong numerical evidence found by us in
the case $3n=3$, it seems reasonable to extend to~(\ref{rteq}) the
validity of what follows:
\begin{itemize}
\item[i)]
  Depending on the choice of $\xbf_0$,
  the limit (\ref{L00}) takes on $3n$ (not
  necessarily distinct) values:
  \[
  \lambda_1(\rz)\geq\lambda_2(\rz)\geq\ldots\geq\lambda_{3n}(\rz).
  \]
\item[ii)]
  If $\xbf_0$ is chosen arbitrarily, the value of the limit~(\ref{L00})
  is $\launo(\rz)$.
\item[iii)]
  Consider the symmetric and positive matrix
  \bb
  W_{\rzs}(t) :=
    \left(\widetilde{U}_{\rzs}(t)\,U_{\rzs}(t)\right)^{1/2t}\,,
  \label{eq19}
  \ee
  where $\widetilde{U}$ denotes the transpose of $U$.
  Denote by $\mu_1(t,\rz)\ge\ldots\ge\mu_{3n}(t,\rz)$ its eigenvalues.
  Let
  \bb
    \lai(t,\rz) := \ln \mu_i(t,\rz), \quad i=1,\ldots,3n.
    \label{eq19bis}
  \ee
  Then,
  \[
    \lim_{t\to+\infty}\lai(t,\rz)=\lai(\rz).\quad\square
  \]
\end{itemize}

The meaning of the LEs is put well in evidence by
the following  geometric argument.
Consider the singular value decomposition~\cite{Atkinson}
of the matrix $U_{\rzs}(t)$, i.e.
\bb
U_{\rzs}(t) L_{\rzs}(t) = G_{\rzs}(t) F_{\rzs}(t),\quad t>0,
\label{eq20}
\ee
where $L$ and $G$ are orthogonal matrices
and $F$ is a diagonal matrix with elements
$\sigma_1(t,\rz) \ge \sigma_2(t,\rz) \ge
\ldots \ge \sigma_{3n}(t,\rz) > 0$.
Take account of the fact that~(\ref{eq20}) implies
\[
\widetilde{L}_{\rzs}(t) W_{\rzs}(t)^{2t} L_{\rzs}(t) = F_{\rzs}^2(t),
\]
so that the columns of $L$ are the orthonormal eigenvectors
$\ebf_1(t,\rz), \ldots, \ebf_{3n}(t,\rz)$ of $W_{\rzs}(t)$ and
\bb
\sigma_i(t,\rz)=\exp[\lai(t,\rz)t], \quad i=1,\ldots,3n.
\label{eq20bis}
\ee
Then,
\bb
  U_{\rzs}(t) \ebf_i(t,\rz) =
     \exp[\lai(t,\rz) t ] \ebf'_i(t,\rz), \quad i=1,\ldots,3n,
\label{eq21}
\ee
where $\ebf'_i(t,\rz)$ are the columns of $G$.
Let now $d\rz$ be an infinitesimal parallelepiped,
with axes \ $\eps_i \ebf_i(t,\rz)$, $i=1,\ldots,3n$, which at $t=0$
is placed at $\rz$.
Since the $\eps_i$ are infinitesimal,
\bbas
  U_{\rzs}(t) \eps_i \ebf_i(t,\rz) \simeq
  \Phi(t,\rz+\eps_i \ebf_i(t,\rz)) - \Phi(t,\rz)&,&\\
  \quad i=1,\ldots,3n&,&
\eeas
and thus, in virtue of~(\ref{eq21}), the phase flow at $t>0$ transforms
$d\rz$ into another infinitesimal parallelepiped
$d\rbf(t)$, which presents a different orientation in $\R^{3n}$
and the $i$th axis, $i=1,\ldots,3n$, grown or shrunk according as the
corresponding $\lai(t,\rz)$ is positive or negative, respectively.
The value of $\lai(t,\rz)$ gives the rate
 at time $t$ of the exponential growing or shrinking of this axis.
The meaning of  $\lai(\rz),\ i=1,\ldots,3n$,
is then obtained accordingly in the limit $t\to+\infty$~\footnote{All the
quantities introduced above, $\lai(\rz)$, $\lai(t,\rz)$, $U_{\rzs}(t)$, etc.,
depend on the initial time $t=0$. But according to footnote 3, we have not
displayed $0$.}.
We can now prove the following property:\\

{\bf Property 4.}
{\sl
Let $d\rz$ and $d\rbf(t)$ be volume elements the same as in Property 2.
Suppose $d\rz$ is a parallepiped. Then,
\bb
  d\rbf(t) =
  d\rz\,\exp\left[\sum_{i=1}^{3n}\lai(t,\rz)t\right],
  \qquad \forall t>0\,.
  \label{eq22}
\ee
}

{\sl Proof.} In virtue of~(\ref{eq21}), each axis
\[
d\xbf_i = \eps^i_1 \ebf_1(t,\rz)+ \ldots + \eps^i_{3n} \ebf_{3n}(t,\rz),
\quad i=1,\ldots,3n,
\]
of $d\rz$ is transformed according to
\bbas
U_{\rzs}(t) d\xbf_i = d\xbf'_i =
    &&\eps^i_1 \exp[\lambda_1(t,\rz) t ]\ebf'_1(t,\rz) + \ldots +\\
    &&\eps^i_{3n} \exp[\lambda_{3n}(t,\rz) t ]\ebf'_{3n}(t,\rz).
\eeas
Taking account of this fact and recalling
that the $3n$-dimensional volume spanned by $3n$ linearly independent
vectors arranged in an $3n\times 3n$ matrix $A$ is $|\det(A)|$, the proof
is completed by a straightforward calculation.$\square$\\

The relationship~(\ref{eq22}) entails two more properties.\\

{\bf Property 5.}
{\sl
\bb
\rho\rtt = \rho_0(\rz)\,\exp\left[ -\sum_{i=1}^{3n}\lai(t,\rz)t\right],
\qquad \forall\,t>0,
\label{eq23}
\ee
where the points $\rbf(t)$ and $\rz$ are the same as in Property 2
and $\rho\rtt=\psi^\star\rtt\psi\rtt $.
}\\

{\bf Property 6.}
{\sl
\bba
\sum_{i=1}^n\,\frac{1}{m_i}\,(\lap_iS\rtt) =
\sum_{i=1}^{3n}[\lai(t,\rz) + \dot{\lambda}_i(t,\rz)t]&,& \nonumber \\
                                     \quad\forall\,t>0&,& \label{eq24}
\eea
where $S$ is the phase of $\psi\rt$ in the polar form~(\ref{eq2bis})
and the points $\rz,\rbf(t)$ are the same as in Property 2.
}\\

The proof of~(\ref{eq23}) is a straightforward consequence of~(\ref{eq22})
and of Property 2.
To prove~(\ref{eq24}), it is enough to differentiate with
respect to $t$ Eq.~(\ref{eq22}) and then to
compare the result with~(\ref{eq16}) and to recall~(\ref{eq3})
and~(\ref{defvi}).
These two latter properties seems to us somewhat intriguing,
for the following reason.
The quantities $\lai(t,\rz)$
are functions of $t$ and hence of the point $\rbf(t)=\Phi(t,\rz)$.
Therefore, they may be looked
upon as ``classical dynamical variables'' of the point pursuing the
the trajectory $\Phi(\cdot,\rz)$.
In virtue of the relationships~(\ref{eq23}) and~(\ref{eq24}), these
variables are strictly correlated to the time-evolution of
$\psi\rt$ along the given trajectory.

\section{Quantum Lyapunov Exponents}

In this section, we give the precise definition of chaotic solution
for the IVP~(\ref{eq9}) and at the same time we give the tool
to measure the corresponding degree of
quantum chaos. We get this tool by considering a properly weighted
average of the LEs of all trajectories in the associated QF.\\

{\bf Definition 1.}
{\sl
Let $\psi\rt$ be the solution of the IVP~(\ref{eq9}).
Recall that $\rho_0(\rbf(0))$ is the corresponding probability
density of points in our statistical ensemble $\Ec$ at $t=0$.
Then, the quantities
\bb
\Lai(0) :=
\int_{\Pc(0)} \lai(\rbf(0)) \rho_0(\rbf(0)) d \rbf(0),
\quad i=1,2,\ldots,3n,
\label{eq25}
\ee
where $\lai(\rbf(0))$ is the $i$th LE of the trajectory of~(\ref{eq10})
starting from $\rbf(0)$ at $t=0$,
are called ``quantum Lyapunov exponents'' (QLE) of the given $\psi\rt$.
For the sake of simplicity hereafter we suppose the usual normalization
\ $N\equiv\int_{\Pc(0)}\rho_0(\rbf(0))d\rbf(0)=1$; \ if $N\neq 1$, we must   
divide by $N$ the r.h.s. of eq.\,(26).}\\

{\bf Property 7.}
{\sl
The QLEs are invariant under the QF, i.e.,
\bbas
  \Lai(0) = \Lai(t) := \int_{\Pc(t)} \lai\rtt \rho\rtt d \rbf(t)&,& \\
                         i=1,2,\ldots,3n, \quad \forall t>0&,&
\eeas
where $\rbf(t)=\Phi(t,\rbf(0))$ and $\lai\rtt$  is the LE of the
trajectory of~(\ref{eq10}) starting from $\rbf(t)$ at $t$.
}\\

{\sl Proof.} As in the proof of Property 2,
think of the QF at $t$ as the coordinate
transformation $\Phi(t,\rbf(0))=\rbf(t)$, from $\Pc(0)$ to $\Pc(t)$.
We find
\bbas
\Lai(t)
&:=& \int_{\Pc(t)} \lai(\rtt)              \rho(\rtt)  d \rbf(t) \\
&=& \int_{\Pc(0)} \lai(\Phi(t,\rbf(0)),t) \rho_0(\rbf(0)) d \rbf(0) \\
&=& \int_{\Pc(0)} \lai(\rbf(0))           \rho_0(\rbf(0)) d \rbf(0) \\
&=:& \Lai(0), \quad i=1,2,\ldots,3n, \quad \forall t>0,
\eeas
where we have employed~(\ref{invar}) and the property
$\lai(\Phi(t,\rbf(0)),t) = \lai(\rbf(0))$
which follows from the very definition of LE.$\square$\\

According to the above property, the QLEs are  parameters whose values are
univocally correlated to peculiar features of a given quantal evolution, i.e.\
the chaotic behaviour of the trajectories in the associated QF.
Therefore, in correspondence with the definition of classical chaos, we can
introduce the following definition:\\

{\bf Definition 2.}
{\sl
A given solution of the IVP~(\ref{eq9})
is called ``chaotic'' if the leading QLE
\ $\Launo$~\footnote{Henceforth, we shall denote the QLEs by $\Lai$ instead of
$\Lai(0)$.} is positive.
The value of $\Launo$ is assumed as a measure of the
corresponding quantum chaos.
}\\

Notice that employing only $\Launo$ in the definition above
corresponds to the fact that the quantity of main interest, to quantify
chaos in a dynamical system, is the leading LE. Another quantity is the
so-called KS entropy~\cite{Hilbor,Benettin3} which requires all
the positive LEs.
In principle, an analogous parameter employing all the positive QLEs,
could be introduced also here, but we do not find it necessary
to discuss this point further in the present paper.

\section{Numerical Results}

Since the leading QLE is a precise indicator of our characterization of 
quantum chaos, most of the present section is devoted to discuss in some 
detail, through a numerical example, how it is possible to calculate it. We 
also show numerical results to illustrate some properties of the QF and the 
complex variety of the underlying trajectories. As is evident from the 
definition~(\ref{eq25}), the computation of $\Launo$ is, in general, very 
difficult. Indeed, once assigned a given solution~\footnote{In this section, 
we call such a solution simply {\em wave-packet}.} $\psi\rt$
of the IVP~(\ref{eq9}), this calculation requires what follows:
\begin{itemize}
\item[1)] Numerical integration of the IVP~(\ref{eq10}),
          associated with the given wave-packet $\psi\rt$.
\item[2)] Calculation of the LE $\lambda_1(\rbf(0))$ for all trajectories
          in the QF relative to $\psi\rt$.
\item[3)] Evaluation of the integral~(\ref{eq25}), in the case $i=1$.
\end{itemize}
As we shall see, each point above presents peculiar numerical problems.
The numerical example shown below is relative to the calculation of
$\Launo$, in the case of the wave-packet
\bba
\psi\rt &=& \frac{1}{\sqrt{3}} \left(
\varphi_{100}(\rbf) \erm^{\frac{i}{2}t} +
\varphi_{200}(\rbf) \erm^{\frac{i}{8}t} +
\varphi_{211}(\rbf) \erm^{\frac{i}{8}t}
\right) \nonumber \\
&=& \rho^{1/2}\rt \erm^{iS\rts},
\label{eq26}
\eea
where $\varphi_{nlm}(\rbf)$ denote eigenfunctions of the hydrogen
atom~\footnote{Throughout this section, atomic units (a.u.) are employed.}.
The chosen wave-function is the most simple superposition of the hydrogen 
atom eigenstates entailing, as we are going to show, the presence of
quantum chaos and non-vanishing QLEs.
As concerns point 1), we employ an adaptive fourth-order
Runge-Kutta method, evaluating numerically the velocity field
$\vbf\rt = \nabf S\rt$ appearing in~(\ref{eq10}).
We find that a considerable portion of the trajectories
are chaotic, i.e.\ $\lambda_1(\rbf(0))>0$.
This fact implies the well known difficulty \cite{Parker}
about the integration of chaotic systems:
numerical errors propagate exponentially and the trajectories
after a certain time, depending on the
accuracy of the algorithm employed, become meaningless.
This phenomenon is amplified if, as in our case (see below),
there is no attractor.
However, rather than in trajectories, we are interested in the
corresponding LEs, and these latter (see Fig.~1) are stable to within $10\%$.

As concerns point 2), we should evaluate only $\lambda_1(\rbf(0))$,
the largest of the three LEs which appear in our example.
However, since we want to check the relationship~(\ref{eq23})
and also for reasons (see below) of numerical control,
actually, we calculate all three LEs.
It is well known \cite{Parker,Benettin2} that, contrarily
to the evaluation of the only leading LE, the calculation of all of them
present numerical difficulties and special techniques are called for.
In the present paper we have  employed the algorithm introduced in
Ref.~\cite{Benettin2} (BGGS algorithm) and
described also in Ref.~\cite{Parker}.

This algorithm is based on formula~(\ref{L00}),
when one is interested only in the evaluation of the leading LE,
otherwise it employs the more general formula
\bba
\launo(\rz) + \ldots + \lambda_p(\rz) &&= \nonumber \\
\lim_{t\to+\infty} \frac{1}{t} \ln {\rm Vol}^p
\{ U_{\rzs}(t) \xbf_1 \ldots &&U_{\rzs}(t) \xbf_p \}, \nonumber \\
\quad 1\leq p\leq 3n,
\label{eq27}
\eea
where $\xbf_1,\ldots,\xbf_p$ are orthonormalized starting vectors
arbitrarily chosen, ${\rm Vol}^p\{\cdot\}$
denotes the volume of the parallelepiped defined by
the vectors $U_{\rzs}(t)\xbf_1,\ldots,U_{\rzs}(t)\xbf_p$
and $\rz$ is the specific point from which the trajectory starts at $t=0$.
The peculiarity of the BGGS algorithm is that of making use of successive
orthonormalization procedures in order to avoid
well-known~\cite{Parker,Benettin2} problems of overflow and
of ``alignment'' of the starting vectors.
With this trick, the BGGS algorithm turns out to be very effective and
preferable to the evaluation of the LEs directly
as logarithms of the eigenvalues of the matrix~(\ref{eq19}).
Notice that the BGGS algorithm yields, in pratice, approximations
$\lai(t,\rz)$, $i=1,\ldots,3n$, as functions of $t$ for the LEs.
In comparison with the analogous quantities~(\ref{eq19bis}),
the BGGS approximations are, as we said, more reliable numerically.
On the other hand, contrarily to~(\ref{eq19bis}),
they depend on the starting vectors $\xbf_1,\ldots,\xbf_p$, and thus,
strictly speaking, they cannot be considered as the ``dynamical variables''
which enter into the relationships~(\ref{eq22})-(\ref{eq24}).
However, in spite of the fact that the quantities~(\ref{eq19bis}) and
the BGGS approximations coincide only in the limit $t\to+\infty$,
the respective sums of all them coincide at each $t$.
This latter property may be seen comparing Eq.~(\ref{eq22}) with
the analogous one
obtained by truncating at $t$ Eq.~(\ref{eq27}) in the case $p=3n$.
In virtue  of this coincidence, we shall employ the quantities $\lai(t,\rz)$
obtained by the BGGS algorithm to check relationship~(\ref{eq23}).
The numerical difficulty of point 2) is that the wave-packet
(\ref{eq26}), nothwithstanding its semplicity, gives rise to
a complex scenario of trajectories.
In fact, we find trajectories whose behaviour passes,
through all the intermediate degrees, from the regular one to
the chaotic one including also the so-called ``intermittency''~\cite{Hilbor}.
In this latter case, the trajectory switches back and
forth between two qualitatively different behaviours, which in our case are
the mildly chaotic one and the chaotic one.
To put in evidence these different regimes, we show
a typical regular trajectory (Fig.~2a),
a typical ``intermittent'' chaotic trajectory (Fig.~2b) and
a typical chaotic trajectory (Fig.~2c),
and for each of them, together with the LE $\lambda_1(t,\rz)$~\footnote{The
point $\rz$ denotes the starting point of each trajectory. We do not need to
specify this point, which, obviously, varies with the trajectory.}
as function of $t$, we display the quantities
\bb
\lambda^{(k)} := \frac{1}{\Delta t} \ln
\frac{ \| U_{\rbfs(t_k)}( t_{k+1}, t_k ) \xbf(t_k) \| }
     { \|                                \xbf(t_k) \| },
\label{eq28}
\ee
where $k=0,1,2,\ldots$, $\rbf(t_k) = \Phi(t_k,\rz)$,
$t_k = k \, \Delta t$ and
$\xbf(t_{k+1}) = U_{\rbfs(t_k)}(t_{k+1},t_k)\xbf(t_k)$.
The symbol $U_{\rbfs(t_k)}(t_{k+1},t_k)$
denotes the flow matrix solution of the
IVP~(\ref{eq18}) with initial condition
at $t=t_k$, \ $U_{\rbfs(t_k)}(t_k,t_k)=I$.
In other words, the $\lambda^{(k)}$ measure the rate of the exponential
divergence or convergence, in the time interval $(t_{k+1},t_k)$,
of the two trajectories: $\Phi(\cdot,\rz)$ (the fiducial trajectory)
and $\Phi(\cdot,\rz+\eps\xbf_0)$, where $\eps$ is infinitesimal and
$\xbf_0$ is an arbitrary starting vector, $\|\xbf_0\|=1$.
Furthermore, recalling the chain rule for the flow matrix (cf.\ point f)
in Sec.~2), the definition~(\ref{L00}) and point ii) in Sec.~3, we have
\[
\lambda_1(t_N,\rz) :=
\frac{1}{t_N} \ln \| U_{\rzs}( t_N ) \xbf_0 \| =
\frac{1}{N}   \sum_{k=0}^{N-1} \lambda^{(k)}.
\]
Strictly speaking, $\lambda_1(t_N,\rz)$ shows a dipendence on
$\xbf_0$, which disappears in the limit $t\to+\infty$.
Here, we do not find it necessary to investigate such a dependence,
because in our case the LEs present a relative error of order 10\%.
The above mentioned complex scenario of trajectories has
forced us to find out a numerical device (see below) capable
of giving a reliable estimation of
\bb
\launo(\rbf(0)) = \lim_{t\to+\infty} \launo(t,\rbf(0)),
\label{eq29}
\ee
over the whole range of the chaotic behaviour.
The point 3) in our example reduces to the evaluation of
\bb
\Launo = \int_{\Rc} \launo(\rbf(0)) \rho_0(\rbf(0)) d\rbf(0),
\label{eq30}
\ee
where $\rho_0(\rbf(0))$ is the probability density in~(\ref{eq26})
at $t=0$, $\rbf(0)$ varies on $\Rc\subset\Pc(0)$ and
$\Rc=\{(x,y,z):-9<x<9,-6<y<6,-8<z<8\}$
is a region such that the density $\rho_0(\rbf(0))$
is negligible outside.
The numerical difficulty here is that the values of
$\launo(\rbf(0))$ are given, as we said, with a relative
error of order 10\%.
For this reason, we have found it convenient to employ an
adaptive Monte-Carlo method and to perform four integration runs
with different accuracy parameters,~\footnote{The accuracy parameters 
changing in the four different runs, are some parameters which rule the 
numerical accuracy of the employed NAG algorithms as the integration procedure 
of the IVP (9) or the numerical squaring. After changing those parameters 
the computed trajectories can vary a lot because of the presence of chaos. 
Notwithstanding the global QLE, coming out from the integration over the 
whole domain $\Rc$, is not sensitively influenced by the variation of the
accuracy parameters (the change in the QLEs results to be less than 10\%).}
to see how much the errors over $\launo(\rbf(0))$ affect the value of 
$\Launo$.

We pass now to describe how the actual evaluation
of~(\ref{eq30}) is carried out.
Chosen an initial condition $\rbf(0)\in\Rc$ by the Monte-Carlo routine,
the integration of the IVP~(\ref{eq10}) and the BGGS algorithm
start in simultaneously.
After a time, say $t_1$, large enough, we apply the numerical
device previously mentioned to estimate the limit~(\ref{eq29}).
Actually, this numerical device consists in two different
convergence criteria.
The first one is conceived to recognize regular trajectories and
it is as follows. We divide the interval
$I_1=[0.4t_1,t_1]$ in five sub-intervals of equal lenght and compute a
linear least squares approximation (LLSA) to $\launo(t,\rbf(0))$ in
each sub-interval.~\footnote{The lower endpoint of the interval is chosen
equal to $0.4t_i$ because we have no interest in studying the behavior of 
$\launo$ at small time; we will instead establish if $\launo$ 
converges to 0 or not, that is, if the trajectory is regular or not.}
If all the LLSAs are decreasing and the minimum of $\launo(t,\rbf(0))$ on 
$I_1$ is smaller than a fixed threshold,~\footnote{The ``fixed threshold'' 
was estimated through an empirical, cut-and-try procedure and results to 
be of the order of the typical asymptotic value (very close to 0 at 
sufficiently large times) of $\launo$ found for trajectories known to 
be regular.} the corresponding trajectory is considered regular and its LE
$\launo(\rbf(0))$ in~(\ref{eq29}) is estimated at zero (Fig. 3).
As concerns the second convergence criterion, conceived for chaotic
trajectories, we compute a LLSA of the function
$\launo(t,\rbf(0))t$ on $I_1$.
If the root-mean-square-error in the LLSA is sufficiently small,
the angular coefficient of the fitting straight line is assumed
as value of $\launo(\rbf(0))$ (Fig. 4).
The integration of the IVP~(\ref{eq10})  and
the BGGS algorithm stop, if one of the two convergence criteria
is satisfied, otherwise they go on up to a certain time $t_2>t_1$
and the two criteria are again applied on the new interval $[0.4t_2,t_2]$,
and so on, until one of the two criteria is met.
Then the procedure starts again with a new initial condition.

In order to evaluate the QLE~(\ref{eq30}) with a relative accuracy of 10\%,
the Monte-Carlo routine has employed about 1700 initial conditions and the
corresponding CPU-time in a double-precision calculation on a workstation,
has been about 60 hours.
As we said, we have performed four such runs, varying the accuracy parameters,
and the correspondent values found for $\Launo$ are the following:
\ $9.048\cdot 10^{-3}$, \ $9.326\cdot 10^{-3}$, \ $8.465\cdot 10^{-3}$,
\ $9.048\cdot 10^{-3}$.

Nothwithstanding the scenario of the trajectories pertaining
to the wave-packet~(\ref{eq26}) is complex, we have found that our
integration domain $\Rc$ may
be subdivided into three sufficiently  disjoint regions, according to
the $\launo(\rbf(0))$ values.
These three regions (Fig. 5) are: the regular one ($\launo(\rbf(0))=0$),
the mildly chaotic one ($0<\launo(\rbf(0))<10^{-2}$)
and the chaotic one ($\launo(\rbf(0)) \ge 10^{-2}$).
On the basis of the values found for $\Launo$ and
considering also that the most chaotic
trajectories have, in our units, LEs of order $0.04$, we can say that the
wave-packet~(\ref{eq26}) presents a good degree of quantum chaos.
Although we have investigated in detail only the wave-packet~(\ref{eq26}),
we have the numerical evidence, according to what observed
in Ref.~\cite{Parmenter} for a different system, that there are
chaotic trajectories in the QFs corresponding
to most wave-packets relative to the hydrogen atom.
Thus, even if the classical counterpart of the hydrogen atom
is not chaotic, our characterization of quantum chaos seems
to spread through the corresponding quantum dynamics.
However, there are exceptions, such as the following:
\begin{itemize}
\item[i)]
Spherically symmetric wave-packets $\psi(r,t)$.
\item[ii)]
Wave-packets of the~kind~$\psi\rt=\sum_k\alpha_k f_k(\rbf)$
$\erm^{iE_kt}$
where $\alpha_k,f_k \in \R$, and the energies $E_k$ are commensurate
\item[iii)] Stationary wave-packets
$\varphi(x,y,z)\erm^{-iEt}$, where
$\varphi(x,y,z)$ $=\varphi^{\star}(x,-y,z)$.
For example,
$\psi(r,\theta,\phi,t)$ $=\erm^{iEt}\,\sum_k\,\alpha_kf_k(r,\theta)$
$\erm^{im_k\phi}$, where $\alpha_k,f_k\in\R$, $m_k=0,\pm 1,\ldots$.
\end{itemize}
The  wave-packets i), ii) always give rise to regular QFs, whereas the
wave-packet iii) only in some situations.
Although the explanation of these facts is not very difficult,
it cannot be given in few words,
and it will be shown within a more general framework elsewhere.
Here, we confine ourselves to note that iii) includes, in particular,
the stationary solutions
$R_{nl}(r)\,Y_{lm}(\theta,\phi)\,\erm^{-iE_nt}$
of the time-dependent Schr\"odinger equation for the hydrogen atom.
In this latter case, all orbits are circles centred on the $z$-axis
and parallel to the $(x,y)$-plane.
Obviously the corresponding $\Launo$ is zero.

Fig.~6 shows the check of the relationship~(\ref{eq23}).
Although this relationship holds, in principle, for any $t>0$, in practice,
its validity is resticted to a
finite time interval $0<t<T$, where $T$ depends on the numerical accuracy.
In our case $T$ is very large (of order $10^6$ a.u.).
In Fig.~6, to put in evidence discrepancies in a short time interval,
we have reduced the accuracy in the calculation of the LEs.

Fig.~7 illustrates that
\bb
\lim_{t\to+\infty} \sum_{i=1}^{3}\lai(t,\rbf(0)) = 0,
\quad \forall \rbf(0) \in \Rc.
\label{eq31}
\ee
In virtue of the relationship~(\ref{eq23}),
limit~(\ref{eq31}) means that
our trajectories do not reach the nodes of the wave-packet~(\ref{eq26})
and that the density $\rho\rtt$ keeps bounded.
Actually (see Fig.~6), the density $\rho\rtt$ along
a given trajectory shows bounded oscillations.
Thus, recalling Property 2, we can also say that our QF,
as concerns the variation of the size of volume elements,
has not any definite character, but it is expected to be
conservative on average.
We have employed sistematically the limit
relationship~(\ref{eq31}) to check the accuracy of the LE calculation.

Finally, Fig.~8 illustrates the fact that the pattern of a chaotic QF
is very sensitive to tiny modifications in $L^2$-norm of the initial
conditions in the IVP~(\ref{eq9}).
To give a clear idea of this phenomenon, Fig.~8 displays only
the components $z(t)$ (as functions of $t$) of the points pursuing
two typical chaotic trajectories, both starting at $t=0$ from
the same initial point, but pertaining to two differnt QFs.
The first QF is relative to the IVP~(\ref{eq9}) with initial
condition $\psi_0(\rbf)$, obtained by setting $t=0$ in~(\ref{eq26}).
The second QF is relative to the IVP~(\ref{eq9}) with initial
condition $\psi_0'(\rbf)$, obtained from $\psi_0(\rbf)$ by a
tiny modification of the coefficients of the corresponding linear
combination, such that $\|\psi_0 - \psi_0'\| = 10^{-4}$.

\section{Spin contributions for the QLE}

Whenener we deal with a spinning particle ---as is the electron of the hydrogen
atom--- we must take into account
also the presence of the intrinsic angular momentum. The presence of the spin
in the Schr\"odinger theory does not affect the stationary energy levels and
the wave-equation to be used is, as usual, Eq.~(\ref{eq2}). Nevertheless, we have
to consider the spin, not only to obtain the allowed stationary states of atoms
(because of the Pauli principle and the ``exchange interaction''), but also to
write the {\em complete} expression of the conserved probability current for
the Schr\"odinger equation. The complete Schr\"odinger current can be
got~\cite{Landau,Hest,Sal,Esp}: \ a) taking the non-relativistic limit of the
Dirac current; \ b) considering the Pauli current in the particular case of
a spin-eigenstate wave-function; \ c) directly from the Schr\"odinger
equation, by considering the privileged direction which the spin introduces in
the space. Let us remember that any solution of the Schr\"odinger equation is
always a spin eigenstate along some space direction, because the Hamiltonian
$-\hbar^2\lap/2m+U$ does not contain any spin term. Then the ``complete'' form
for Schr\"odinger wave-function for a spinning particle actually is
\ $\sqrt{\rho}\,\erm^{iS/\hbar}\chi$ \ where
$\chi$ is a {\em constant} 2-components Pauli spinor.
Following Landau~\cite{Landau} we can write the conserved Schr\"odinger current
as follows:
\bb
\jbf = \frac{\rho}{m}\,\vecna S + \frac{\hbar}{2m}\vecna\rho\times\sbf\,.
\label{current}
\ee
where $\sbf$ is the constant unit versor in the direction of the spin:
\ $\sbf\equiv\chi^\dagger\sigbf\chi/\chi^\dagger\chi$, \ $\sigbf$ being
the usual Pauli 2$\times$2 vector matrix.~\footnote{Let us incidently remark that,
by contrast with the Schr\"odinger-case, the most general wave-function
solution of the Pauli equation must be written in
the non-decomposed form $\chi\rt$, where the two spinor components vary in
space and in time independently eachother. This happens since the spin is not
anylonger conserved, e.g. because of its coupling $-e\sbf\cdot\Hbf/2m$ with some external
magnetic field. Respect to the Schr\"odinger-case, the spin term in the
conserved current cannot be anylonger written as $\vecna\rho\times\sbf/2m$ but
only as
$\vecna\times\left[\chi\rt^\dagger\sigbf\chi\rt\right]/2m$~\cite{Landau,Sal}.}
When the polarization direction is chosen
parallel to the $z-$axis, i.e., \ $\sbf=(0;\,0;\,s_z), \ s_z=\pm1$, \ we have
for the spin-part of the current:
\bb
\frac{\hbar}{2m}\vecna\rho\times\sbf
\ug s_z\,(\vecna_y\rho;\;-\vecna_x\rho;\;0)\,.
\ee
The current in Eq.~(\ref{current}) appears as a sum of the translational,
``newtonian'', part \ $\rho\vecna S/m$ \ which, at the
classical limit ($\hbar\longrightarrow 0$), is parallel to the classical
impulse (equal to $\hbar$ times the gradient of the action); and of a
``non-newtonian'', rotational part due to the spin, $\vecna\rho\times\sbf/2m$,
which vanishes only at the classical limit, i.e. for spinless bodies, {\em but
not in the non-relativistic limit}, i.e. for a small momentum $\imp$.
\ Let us stress that for {\em any solution of the Schrodinger equation the
spin part of the current is conserved} because the curl of a divergence
identically vanishes:
$$
\vecna\cdot[\vecna\rho\times\sbf] = \vecna\cdot[\vecna\times(\rho\sbf)] = 0\,;
$$
as a consequence, if $\rho\vecna S/m$ is conserved, quantity $\rho\vecna S/m +
\hbar\vecna\times(\rho\sbf)/2m$ is conserved as well:
$$
\pa_t\rho + \vecna\cdot\left(\frac{\rho}{m}\vecna S\right) = \pa_t\rho +
\vecna\cdot\left(\frac{\rho}{m}\vecna S + \frac{\hbar}{2m}\vecna\rho\times\sbf\right) = 0\,.
$$
Let us also remark that {\em the spin part of the conserved current results to
be of the same order of the scalar-newtonian term}; actually, in the case of
{\em real} wave-functions, the spin part is the non-vanishing one in the
probability current.~\footnote{As is known, the eigenfunctions corresponding to
non-degenerate energy eigenvalues are always real: in fact, the phase $S$ is
uniform and
equal to a constant (which, for the ``global phase gauge invariance", may be
assumed equal to zero). We refer ourselves to the $l=0,\,n\geq1$ stationary
states of a particle inside a well (or a box) and of the hydrogen atom, as
well as to the $m=0,\,n\geq 0$ stationary states of a particle in a uniform
magnetic field. Even for some systems with degenerate energy
levels, as, for instance, the spherical harmonic oscillator, the eigenfunctions
are real.
For real wave-functions, neglecting the spin, we should have $\jbf=\vecna S=0$
everywhere. As it was first remarked by Einstein and Perrin and
by de Broglie,~\cite{DeB} such a result seems to be in
contrast, from a classical point of view, with the non-vanishing of the energy
eigenvalue for those stationary states. But if we consider the spin component
of the current ---which depends on the gradient of $\rho$, and not on the
gradient of the phase--- we have a rotational motion around the spin
polarization direction, and the quantum ``zero-point'' energy arises as a
kinetic-like energy.}
Let us finally recall that also in Bohm Mechanics the inclusion of the spin part in
the non-relativistic velocity field is requested for spinning
systems~\cite{Deo,Holland}.
For a $n-$particles system Eq.~(\ref{eq3}) generalizes as follows in the presence of
spin:
\bb
\jbf_i\rt := \frac{\rho\rt}{m_i}\,\vecna_iS\rt
+ \frac{\hbar}{2m_i}\vecna_i\rho\rt\times\sbf_i\,.
\ee
while the velocity field $\jbf_i\rt/\rho$, which for a spinless particle
consists of the only term $\vecna_iS\rt/m_i$, now becomes
\bb
\vbf_i\rt := \frac{1}{m_i}\,\vecna_iS\rt
+ \frac{\hbar}{2m_i\rho\rt}\vecna_i\rho\rt\times\sbf_i\,.
\ee
As it is easy to check, all the properties of the QF proved in Sec.~3 still
equally hold in the presence of spin. We have performed several numerical
calculations of the chaotic quantities for the hydrogen atom considering the
spin: that is, by integrating the above ``complete'' velocity field.
We get results quite similar to the ones obtained in the absence of spin
and discussed in the previous section, but the spin part of the quantum flow
seems to increase slightly the chaos with respect to the spinless case.
As an example, for the wave-packet
\bba
\psi\rt &=& \frac{1}{\sqrt{3}} \left(
\varphi_{100}(\rbf) \erm^{\frac{i}{2}t} +
\varphi_{210}(\rbf) \erm^{\frac{i}{8}t} +
\varphi_{211}(\rbf) \erm^{\frac{i}{8}t}
\right)\,,
\eea
with spin projection $s_z=+1$, we have obtained (after four runs, varying
the accuracy parameters, etc.) the following value for the QLE: \
$\Launo$ = \ $1.022\cdot 10^{-2}$.

\section{Concluding remarks}

In this paper, we have proposed a precise definition of chaos in quantum
dynamics.
This definition is based only on basic concepts and tools of the theory of
chaos in dynamical systems and, contrarily to previous papers,
it does require any new interpretation of quantum mechanics.
The characterization of chaos in quantum dynamics introduced by us
may be looked upon, in a sense, as the direct analogue of classical
chaos and thus, it attempts to reduce the gap
between classical chaos and quantum chaos according to the definition
in most current literature.
The present work stimulates  further investigations  mainly
on two fundamental problems.
The first one concerns the correlation with classical chaos. We expect that,
if we adopt the
criterion for the classical limit described in Ref.~\cite{Holland},
quantum chaos here introduced matches with classical  chaos.
The other problem is that of  seeing if the values of the QLEs may
be related to observable quantal quantities.
We want to investigate these two problems in a next paper and work
in this direction is already in progress.

\acknowledgments
The authors would like to thank G.G.N. Angilella, M. Baldo, M. Baranger, S.P.
Goldman, V. Latora, F. Raciti, A. Rapisarda and E. Recami for many helpful
discussions.
This work has been supported by INFN, Sezione di Catania, and by MURST.


\newpage

\begin{figure}
\centering
\includegraphics[scale=0.4,angle=-90]{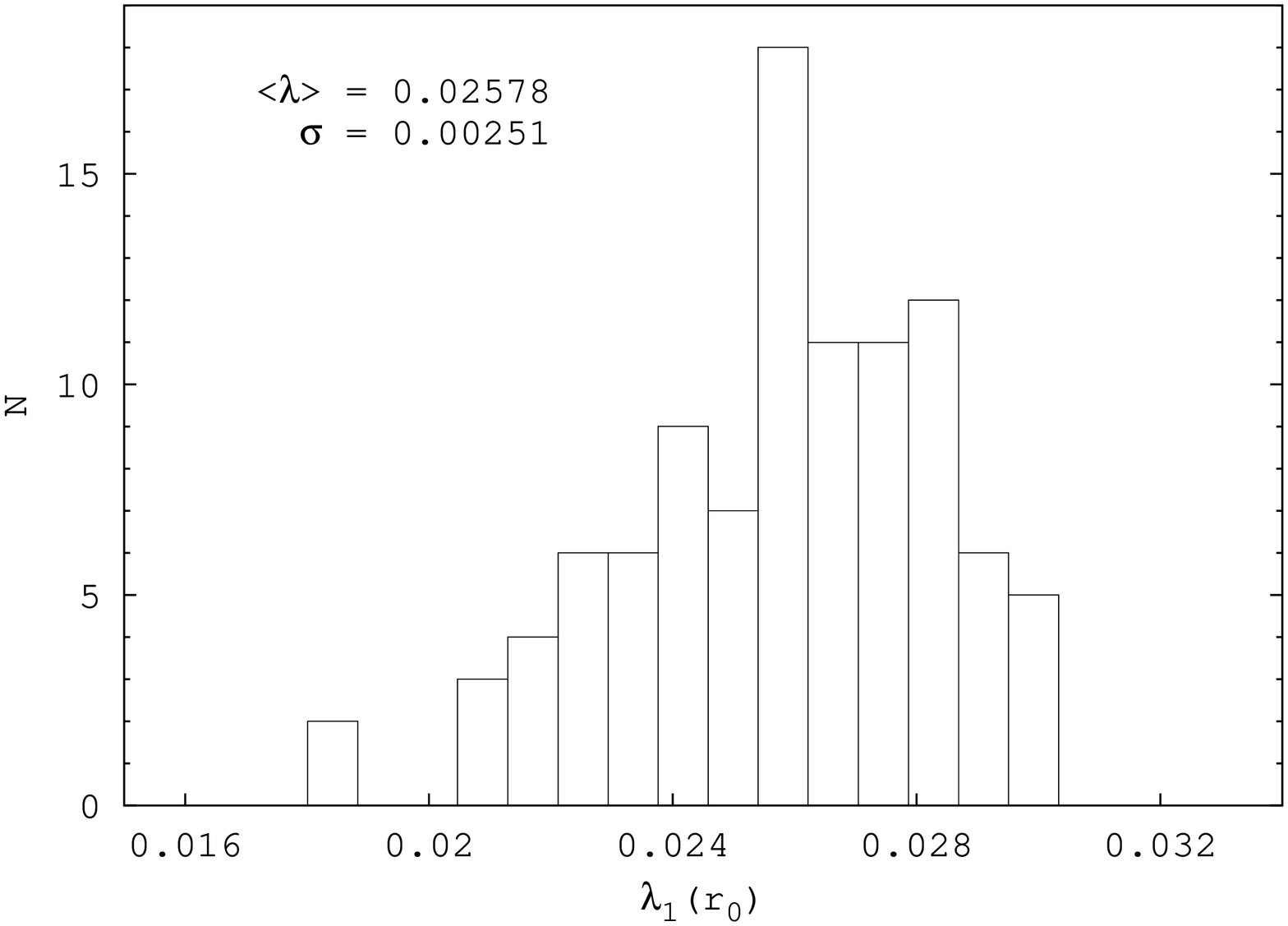}    
\caption{ 
Histogram relative to the distribution of the values found for
$\launo(\rz)$ over 100 trajectories,
all starting from the same initial point $\rz$ but
evaluated with slightly different accuracy.
$N$ is the number of trajectories,
$\langle \lambda \rangle$ is the expectation value
and $\sigma$ is the standard deviation.
}
\label{fig1}
\end{figure}


\begin{figure}
\centering
\includegraphics[scale=0.25,angle=-90]{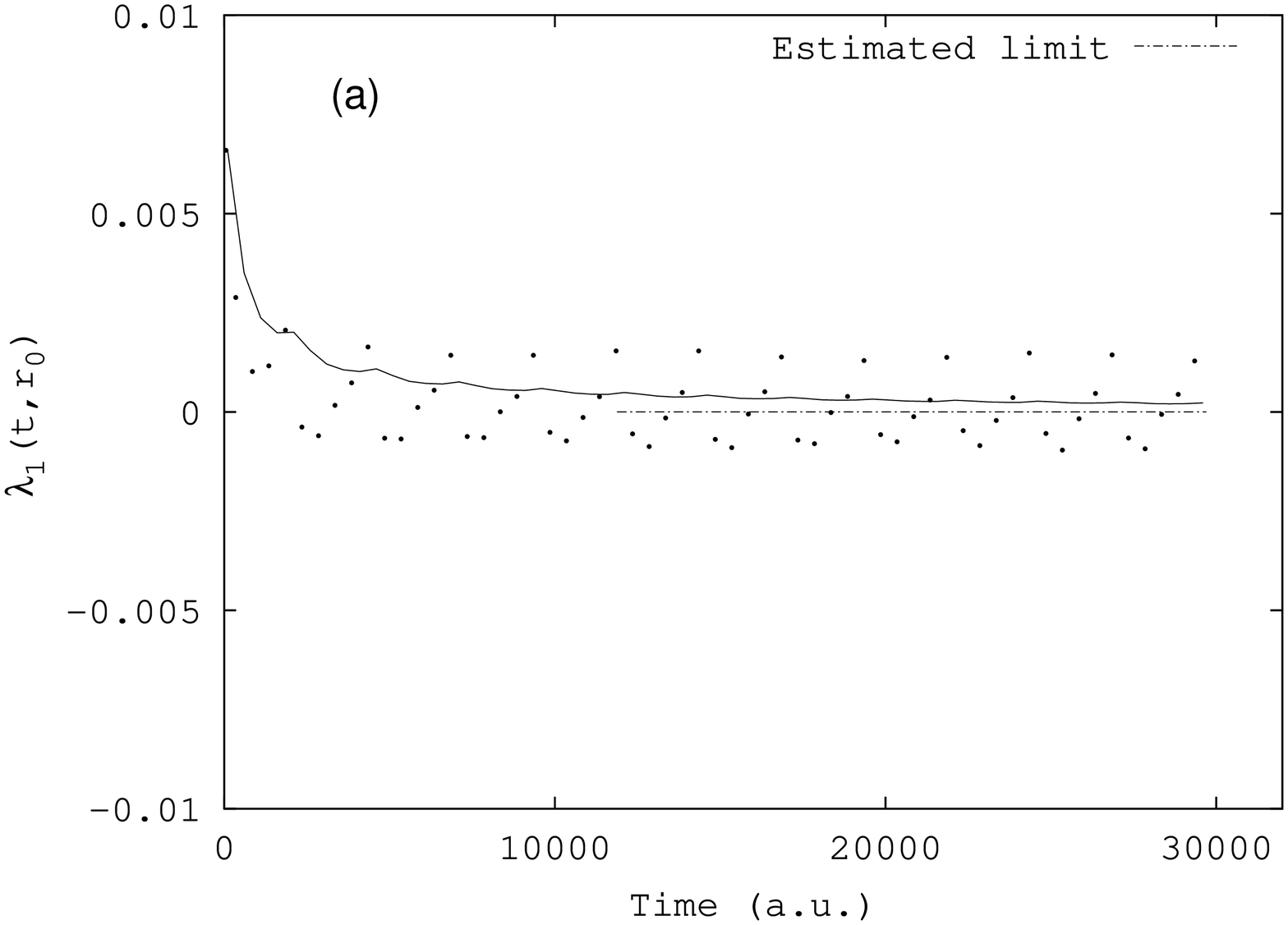}    
\centering
\includegraphics[scale=0.25,angle=-90]{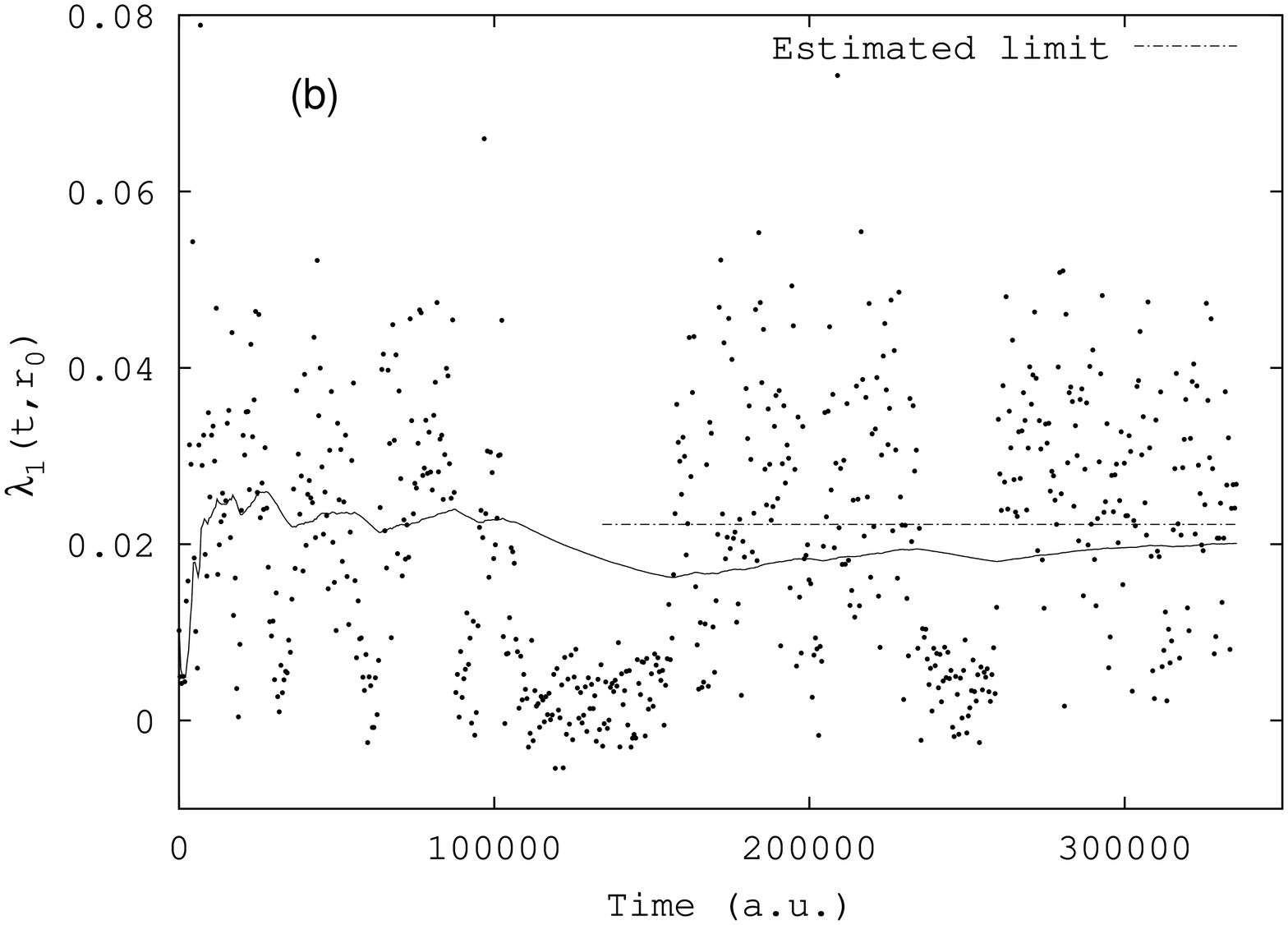}    
\centering
\includegraphics[scale=0.25,angle=-90]{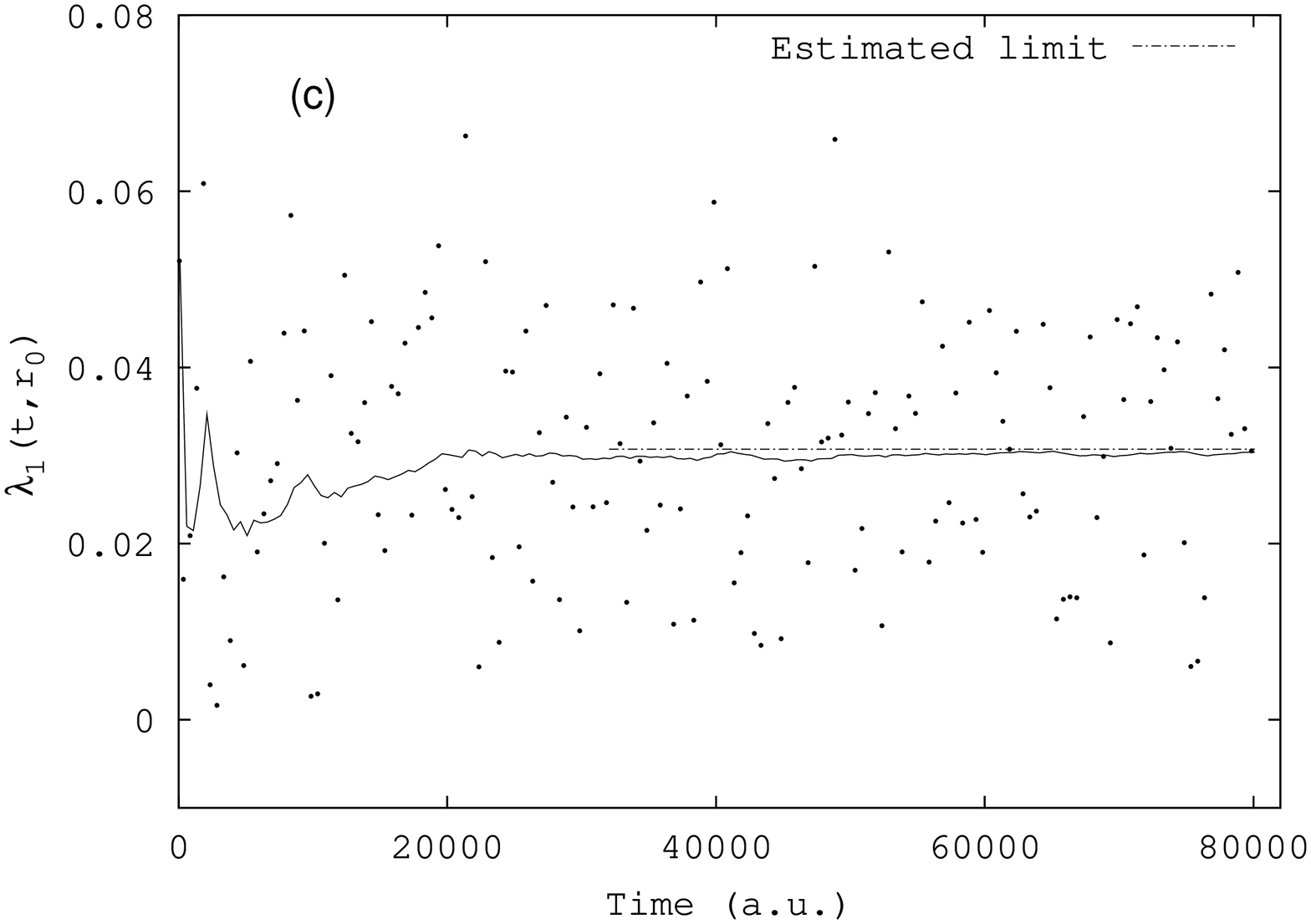}    
\caption{ 
Figures (a)-(c) are relative, respectively, to a typical
(regular, ``intermittent'' chaotic, chaotic) trajectory.
The solid curve represents the LE \ $\launo(t,\rz)$ as function
of $t$ and the dots represent quantities (29).
}
\label{fig2a-c}
\end{figure}

\newpage


\begin{figure}
\centering
\includegraphics[scale=0.4,angle=-90]{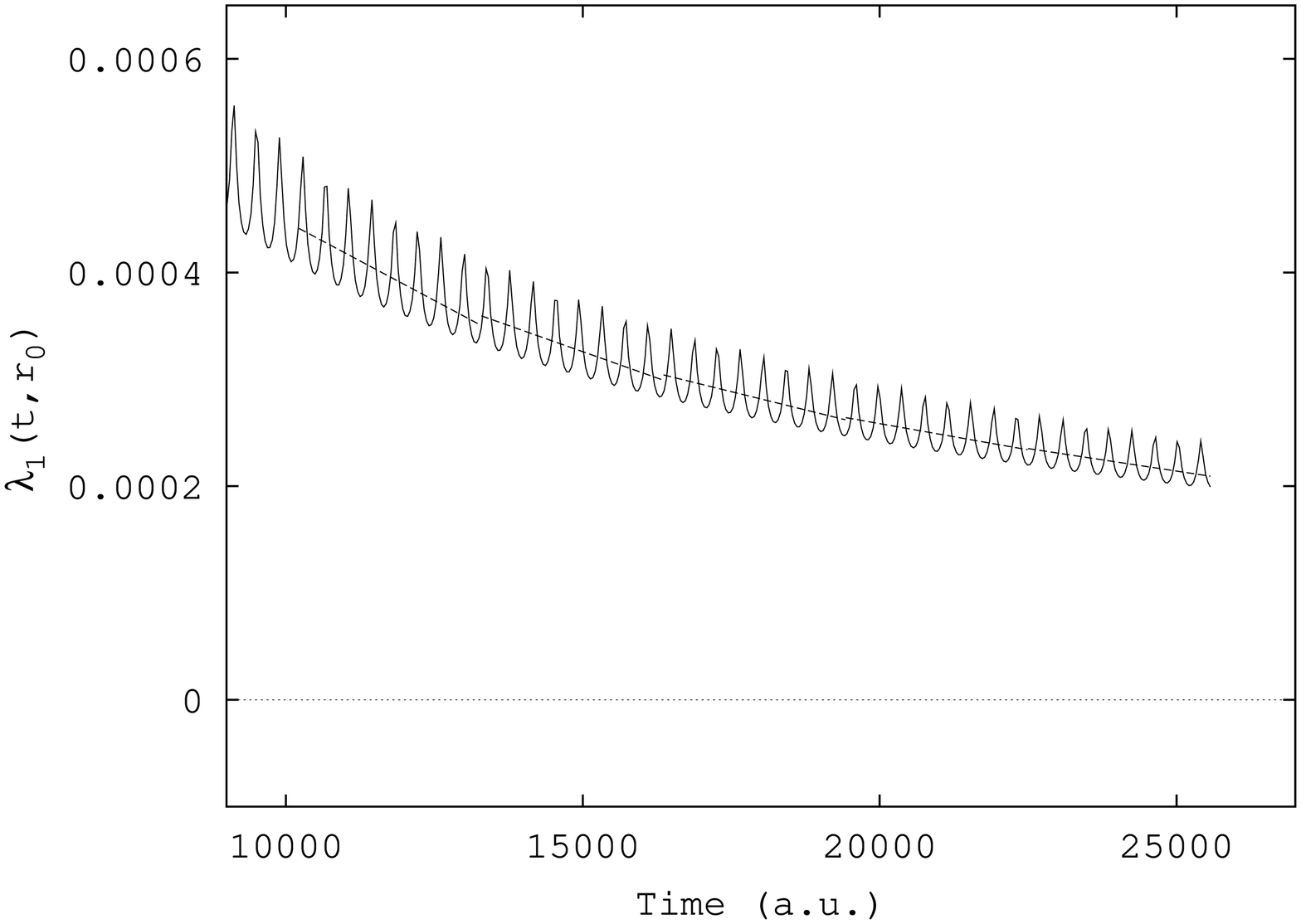}    
\caption{ 
Convergence criterion for a typical regular trajectory starting from $\rz$.
The solid curve represents the LE \ $\launo(t,\rz)$ as a function of $t$.
The dashed straigth lines represent the LLSAs.
}
\label{fig3}
\end{figure}


\begin{figure}
\centering
\includegraphics[scale=0.4,angle=-90]{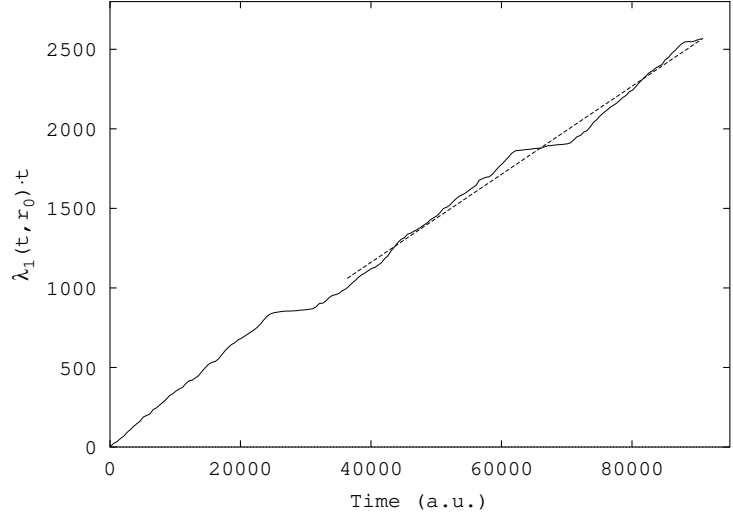}    
\caption{ 
Convergence criterion for a typical chaotic trajectory starting from $\rz$.
The solid curve represents $\launo(t,\rz)t$ as a function of $t$.
The dashed straight line represent the LLSA.
}
\label{fig4}
\end{figure}


\begin{figure}
\centering
\includegraphics[scale=0.4,angle=-90]{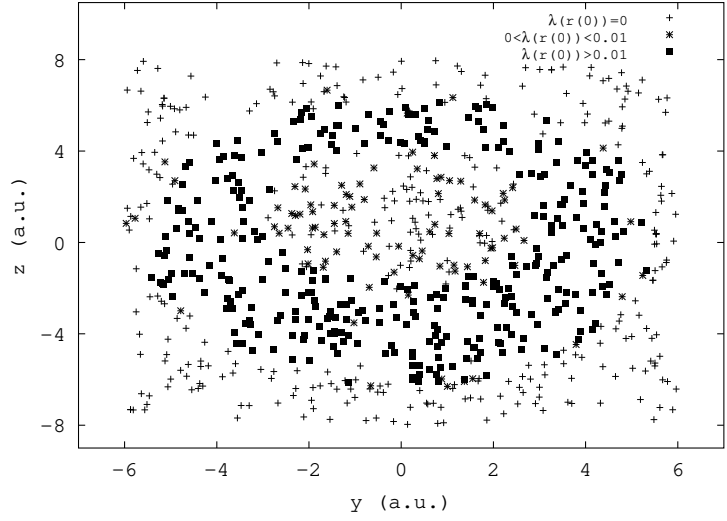}    
\caption{ 
Projection onto the  $(y,z)$-plane of all the initial conditions
in the region $\{(x,y,z)\in\Rc:-0.5<x<0.5\}$ and relative to
the trajectories evaluated in the four runs to calculate $\Lambda_1$.
Three regions of different degree of chaos are put in evidence.
}
\label{fig5}
\end{figure}


\begin{figure}
\centering
\includegraphics[scale=0.4,angle=-90]{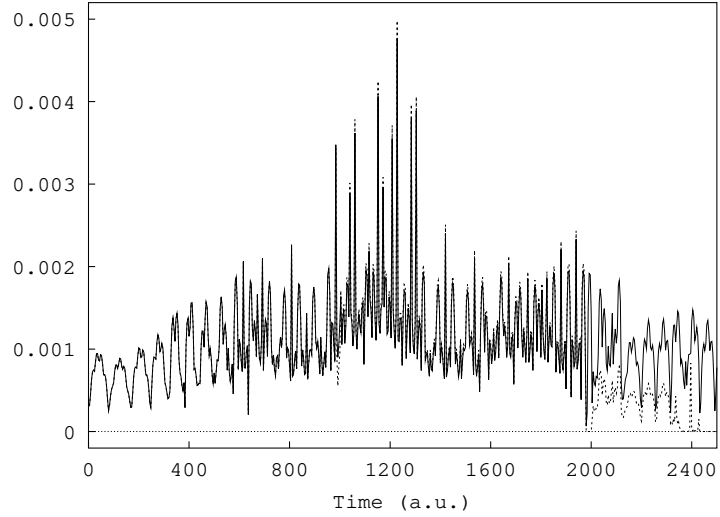}    
\caption{ 
Check of Eq.~(\ref{eq23}).
The solid curve represents the probability density $\rho\rtt$ as given by
the wave-packet~(\ref{eq26}) along a typical chaotic trajectory.
The dashed curve represents $\rho\rtt$ along the same trajectory
but evaluated through~(\ref{eq23}).
}
\label{fig6}
\end{figure}


\begin{figure}
\centering
\includegraphics[scale=0.4,angle=-90]{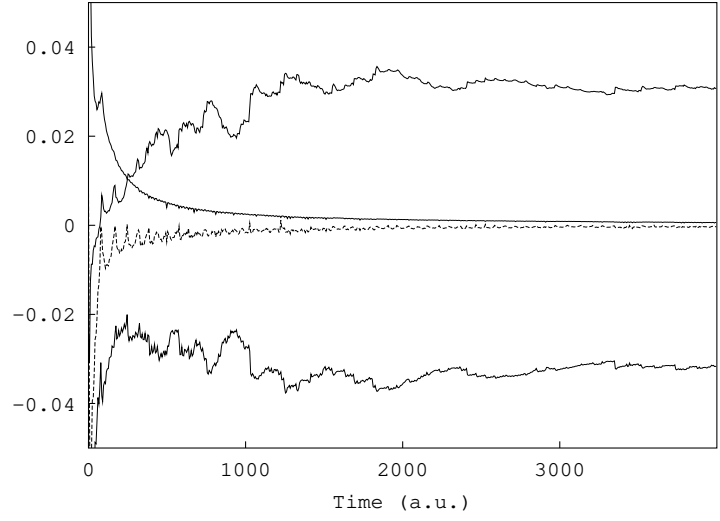}    
\caption{ 
Check of the limit relationship~(\ref{eq31}).
The solid curves represent the three LEs: \
$\launo(t,\rz),\lambda_2(t,\rz),\lambda_3(t,\rz)$ as functions of $t$,
of a typical chaotic trajectory starting from $\rz$.
The dashed curve represents $\sum_{i=1}^3\lambda_i(t,\rz)$.
}
\label{fig7}
\end{figure}


\begin{figure}
\centering
\includegraphics[scale=0.4,angle=-90]{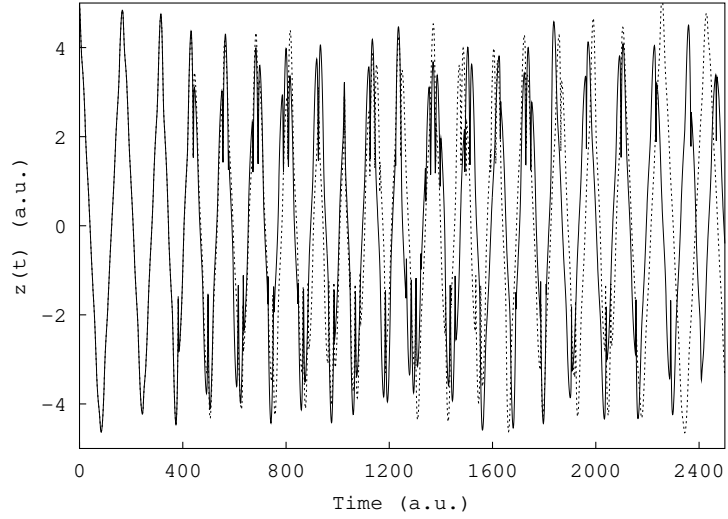}    
\caption{ 
Dependence of the QF pattern on the initial condition
in the IVP~(\ref{eq9}).
}
\label{fig8}
\end{figure}


\end{document}